# Single mode laser in the telecom range by deterministic amplification of the topological interface mode

Markus Scherrer[1], Chang-Won Lee[2], Heinz Schmid[1], Kirsten E. Moselund[3,4]

mas@zurich.ibm.com, cwlee42@hanbat.ac.kr, sih@zurich.ibm.com, kirsten.moselund@psi.ch

[1] IBM Research Europe – Zurich, 8803 Rüschlikon, Switzerland

[2] Institute of Advanced Optics and Photonics, Hanbat National University, Daejeon, South Korea

[3] Laboratory of Nano and Quantum Technologies (LNQ), Paul Scherrer Institut (PSI), 5232 Villigen, Switzerland

[4] Integrated Nanoscale Photonics and Optoelectronics Laboratory (INPhO), EPFL, 1015 Lausanne, Switzerland

* Correspondence: kirsten.moselund@psi.ch, +41 56 310 34 15

**ABSTRACT**

Photonic integrated circuits are paving the way for novel on-chip functionalities with diverse applications in communication, computing, and beyond. The integration of on-chip light sources, especially single-mode lasers, is crucial for advancing those photonic chips to their full potential. Recently, novel concepts involving topological designs introduced a variety of options for tuning device properties such as the desired single mode emission. Here we introduce a novel cavity design that allows to amplify the topological interface mode by deterministic placement of gain material within the topological lattice. The proposed design is experimentally implemented by a selective epitaxy process resulting in Si and InGaAs nanorods embedded within the same topological lattice. This results in the first demonstration of a single-mode laser in the telecom band using the concept of amplified topological modes.

## INTRODUCTION

The rapid advancement of photonic integrated circuits (PICs) is currently driving innovation in diverse fields of modern technology, ranging from data communication[1–4], quantum computing[5–7], sensing[8,9] to automotive applications such as LIDAR[10]. Despite remarkable progress in component integration, the absence of an integrated on-chip light source remains a crucial limitation on silicon (Si)-based platforms[11–13]. While conventional bulk III-V semiconductor lasers have served as reliable workhorses in photonics, the need to seamlessly integrate these lasers onto a photonic chip demands a reimagining of their design. Scaling down lasers not only facilitates their integration but also offers the potential for improved energy efficiency, enabling high-speed data transmission while minimizing power consumption[12,14–17]. Crucially, these nano-scaled lasers must preserve single-mode operation, a key requirement for many applications.

Here we introduce a novel concept that leverages the principles of topological photonics to achieve on-chip, down-scaled, and single-mode lasers. Topological photonic systems are known to exhibit intriguing properties, such as robustness against disorder and scattering, enabling the stable formation and propagation of edge states within the photonic bandgap[18–21]. Recently, new concepts based on non-Hermitian Hamiltonian systems have been developed that harness the combination of active materials with added loss. This can be used to implement complex Hamiltonians[22–24] or study parity-time (PT) symmetric systems[25–27] and to achieve inherently single-mode photonic cavities. Previous studies[25,28] have shown that deterministic placement of loss may be used to selectively dampen trivial modes, such that the topological edge mode prevails. Even though the integration of lossy materials may be necessary to realize a PT-symmetric Hamiltonian, they will typically lead to significant absorption, and it is more desirable to reduce them as much as possible.

In this work, we propose a path to enhance the topological mode without adding loss to the system and experimentally demonstrate single mode lasing from such a device. The proposed solution is based on selectively addressing the topological interface mode through the placement of gain material side-by-side in the otherwise dielectric topological lattice. Experimentally, this is implemented on a SOI platform using selective epitaxy to form a lattice made from interspersed silicon and indium gallium arsenide (InGaAs) nanorods.

## RESULTS

### Design of the topological lattice for single-mode emission

The design presented in fig. 1(a) shows the photonic cavity to achieve single mode emission. Its underlying lattice structure is inspired by the Su-Schrieffer-Heeger (SSH) model, which describes a dimerized lattice formed through alternating the bond strength between individual lattice sites. A unit cell in this lattice thus contains two separate lattice sites with a different spacing than between the two sites of neighboring unit cells. In our implementation, this lattice is made up of nanorods with a high refractive index $n$ that are embedded in low $n$ oxide for dielectric confinement and placed at alternating distance to each other to achieve two different bond strengths. We choose this embodiment because it allows for small footprint devices and can be implemented with our fabrication technique as detailed below.

Depending on the choice of unit cell for the lattice, i.e. which site is at the termination of the finite lattice, the topological invariant of the lattice, the so-called Zak phase, is either 0 or 1, which can be referred to as the trivial or topological case, respectively[29,30]. The SSH topological lattice is known to show interesting features, namely edge modes located at the end site[28,31,32]. When bringing together two lattices as depicted in fig. 1 a, the central defect could be regarded as an edge shared by both half-lattices. We emphasize this by choosing a symmetric coloring scheme for the two half-lattices. The two nanorods can be of the same or of different materials; for simulations we assume the same refractive index $n = 3.5$.

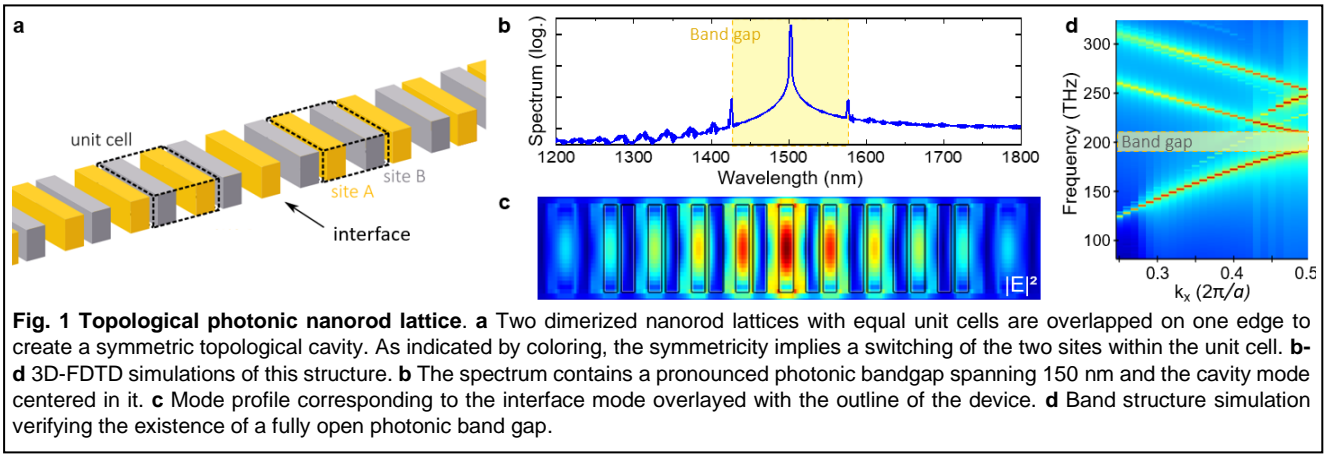

**Fig. 1 Topological photonic nanorod lattice**. **a** Two dimerized nanorod lattices with equal unit cells are overlapped on one edge to create a symmetric topological cavity. As indicated by coloring, the symmetricity implies a switching of the two sites within the unit cell. **b-d** 3D-FDTD simulations of this structure. **b** The spectrum contains a pronounced photonic bandgap spanning 150 nm and the cavity mode centered in it. **c** Mode profile corresponding to the interface mode overlayed with the outline of the device. **d** Band structure simulation verifying the existence of a fully open photonic band gap.

The photonic mode supported at the topological interface (TI) is located – spatially – centered around this interface and – spectrally – in the center of the photonic bandgap. 3D finite-difference time domain (FDTD) simulations (see Methods) confirm this for the given nanorod geometry as shown in fig. 1 b and c. Additionally to this TI mode with an intrinsic quality factor on the order of 20k, the photonic band edges appear in the spectrum as smaller peaks on both sides of the central peak, forming a photonic band gap 150 nm wide. In the band structure presented in fig. 1 d, these two band edges correspond to values at the border of the 1$^{st}$ Brillouin zone.

For our purpose of achieving a single mode laser, the main property of the topological mode which we want to harness here is its distribution in space. As can be seen in fig. 1 c, the electromagnetic field intensity localized only on every other lattice site with the highest intensity at the symmetry center. The simulation further shows that all other modes are more evenly distributed between the two lattice sites and either side of the interface.

This distinct difference in distribution allows the selective amplification of the TI mode by integrating gain material at positions with a high field intensity for this mode. In contrast, the other modes will have a much lower overlap with the gain material at these positions, leading to their suppression and resulting in a high contrast single mode emission spectrum.

**Device implementation**

A key requirement of the device design is the precise placement of two different materials, passive and active, next to each other. Using an SOI platform naturally defines Si as first, passive, material. As gain material a direct bandgap III-V semiconductor is ideally suited, whereby we choose InGaAs for its emission wavelength in the desired telecom range. The co-integration of these two materials is made possible through template-assisted selective epitaxy (TASE), a self-aligned monolithic integration method for III-V semiconductors on Si[33].

Using this technique, we have previously demonstrated high-speed monolithic detectors[34,35] as well as the first photonic crystals emitters showing resonant emission[36].

The epitaxy process to introduce III-V gain material into the desired lattice co-planar with the passive Si nanorods is illustrated in fig. 2 using a single nanorod. It relies on the formation of a hollow silicon dioxide template with a small nucleation seed inside this template for defining the shape and position of the III-V material. During metal-organic chemical vapor deposition (MOCVD) growth, the III-V semiconductors will fill up this template and take on its exact shape. The template itself is defined in the same step as the Si nanorods, it relies on Si as the sacrificial material, thus the two parts are inherently aligned to each other.

A false-colored scanning electron microscope (SEM) top-view image of a final device is displayed in fig. 2 b and highlights the good transfer from device design to fabricated device by TASE. In the center of the image, the Si and III-V nanorods making up the topological SSH chain are visible, whereby they are embedded within the SiO$_2$ layer surrounding the entire structure. The crystal facets on the bottom and top end of the nanorods are angled, since both the Si etch-back and the InGaAs growth, respectively, end up with predominantly {111} facets. For this device, the Si seed extensions were removed to make it more symmetric, but in both our simulations and experiments this does not significantly impact device performance. The following results were achieved on structures which still had the Si seed attached to the InGaAs nanowire.

**Optical characterization**

Fabricated devices are investigated by micro-photoluminescence (µ-PL) spectroscopy (for setup details, see Methods) at a temperature of 100 K. The active material is excited above its bandgap using a 1100 nm pulsed pump laser and emission is collected from the top through a 100x objective. Fig. 3 a shows the power dependent spectrum from a device with a TI mode at 1470 nm. This cavity mode appears already at low powers, shifting slightly in wavelength before this single peak becomes clearly dominant with higher pump powers.

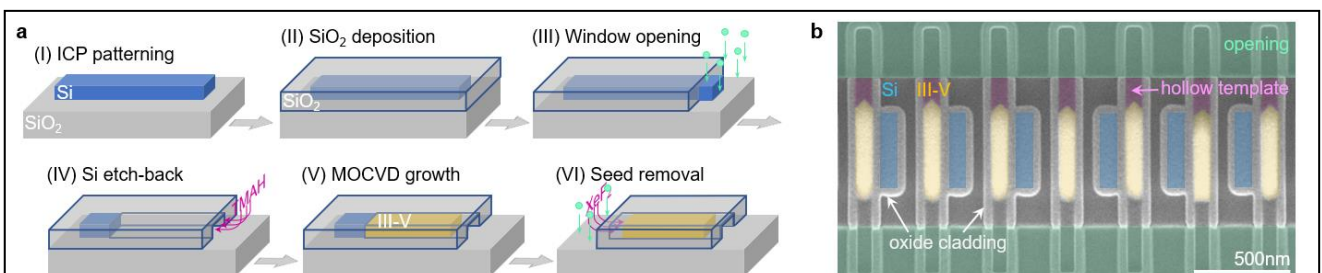

**Fig. 2 Fabrication of the hybrid III-V/Si cavity. a** Process flow for template-assisted selective epitaxy. The structure is patterned by Si dry etching (I), then encapsulated in SiO$_2$ (II). The template is created through locally removing the oxide on one side of the Si structure (III) and then selectively etching out the Si (IV). The resulting hollow template is filled with III-Vs by MOCVD (V). Steps II-IV can be repeated to remove the Si seed on the other side (VI). **b** Scanning electron microscope image of the fabricated structure. The III-V nanorods are visible inside the SiO$_2$ template, alternating with Si nanorods, both of which covered by oxide. Additional features at both ends of each III-V nanorod result from the regrowth procedure: The sacrificial nanorods were designed longer such that they can be accessed on both sides for the etch-back, their outline remains visible as the sidewall of the now empty SiO$_2$ template.

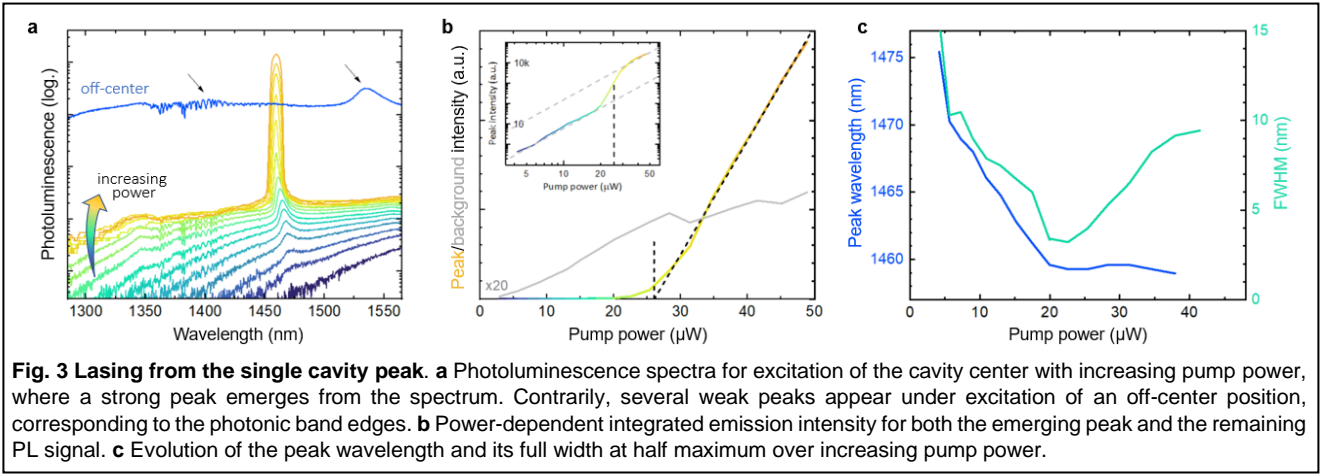

**Fig. 3 Lasing from the single cavity peak. a** Photoluminescence spectra for excitation of the cavity center with increasing pump power, where a strong peak emerges from the spectrum. Contrarily, several weak peaks appear under excitation of an off-center position, corresponding to the photonic band edges. **b** Power-dependent integrated emission intensity for both the emerging peak and the remaining PL signal. **c** Evolution of the peak wavelength and its full width at half maximum over increasing pump power.

We note that this mode is only present when exciting the center of the device with a margin comparable to the width of the pump beam. When instead exciting the outer parts of the device, several weak peaks are visible, which correspond to the location of the photonic band edges (blue curve in Fig. 3 a) of the lattice. We can thereby confirm that the measured mode arises from the topological interface located at the device center.

A clear threshold behavior of the topological cavity mode is observed with increasing pump power - this translates into a kink in the linear light-in light-out (LL) curve characteristic for lasing. The LL curve is shown in fig. 3 b, whereby we separate the contribution of the interface mode from the PL background and integrate the respective counts to get the PL intensity. A threshold of 26 μW is determined from the x-axis intersection of the interpolated line as depicted by the dashed line. Above threshold, the PL background gets clamped as most energy provided by the pump goes into the lasing mode. The inset shows the same data as log-log plot, here the typical transitions in its slope are visible, going from a spontaneous emission regime over amplified spontaneous emission to lasing.

Characteristic differences between the spontaneous and stimulated emission regime, i.e., below and above the lasing threshold can be found also in wavelength and linewidth (i.e. full width at half maximum, FWHM) of the emission peak, as seen in fig. 3 c. Below threshold, we observe a blue shift of the emission wavelength that we attribute to free carrier plasma dispersion effects. This is something typically observed for similarly sized devices[25,37,38]. At the same time, the FWHM of the emission peak narrows significantly with its lowest value of 3 nm at threshold, where it becomes limited by the peak jitter during each individual pulse. Above threshold, the lasing wavelength remains stable under further increase of pump power because the free carrier plasma dispersion effect becomes less pronounced as the spontaneous emission background gets clamped. Additionally, the redshift due of heating of the device compensates the previous blueshift.

We performed time-resolved PL measurements to further evaluate how the carrier lifetime is influenced by increasing the pump power, as shown in fig. 4 a. While the curves below threshold show the same emission behavior with a lifetime of around 37 ps, we observe a reduction of lifetime to 20 ps above threshold. This can be explained by the increasing dominance of the stimulated emission of photons into the TI peak and thus a further proof of lasing, which we demonstrate here for the first time on this platform. Note that the intensity of collected light in the objective above the sample is due to scattering only, as the main propagation direction within the cavity lies along the length of the device.

So far, we have selectively addressed and enhanced the TI mode by inserting InGaAs nanorods in positions with high mode overlap. In a control experiment presented in fig. 5, we introduce the same quantity of gain material into the cavity, but instead place it on those sites with a low intensity of the TI mode. Thus, the position of the III-V nanorods in device B presented in fig. 5 a is the exact opposite to the previous structure, here shown again as device A for direct comparison. As expected, the corresponding photoluminescence spectrum in fig. 5 b does not present the pronounced lasing peak visible in device A. Instead, several weak peaks related to the bulk modes appear.

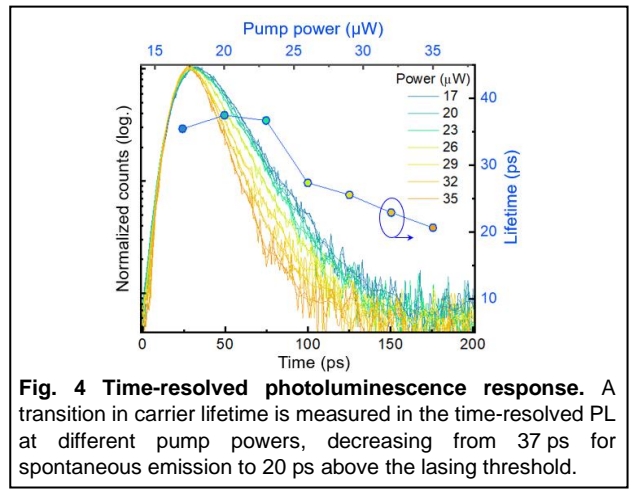

**Fig. 4 Time-resolved photoluminescence response.** A transition in carrier lifetime is measured in the time-resolved PL at different pump powers, decreasing from 37 ps for spontaneous emission to 20 ps above the lasing threshold.

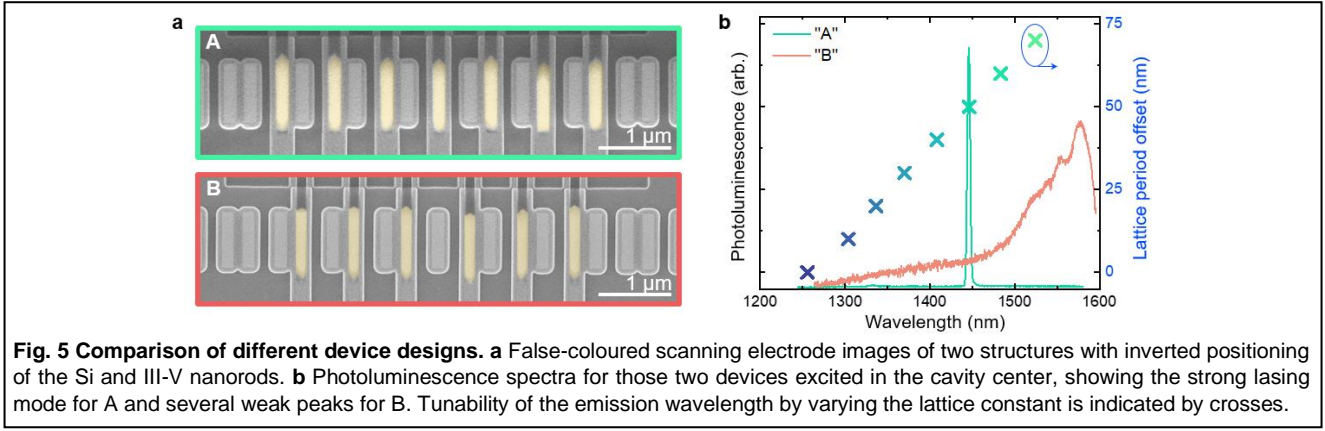

**Fig. 5 Comparison of different device designs. a** False-coloured scanning electrode images of two structures with inverted positioning of the Si and III-V nanorods. **b** Photoluminescence spectra for those two devices excited in the cavity center, showing the strong lasing mode for A and several weak peaks for B. Tunability of the emission wavelength by varying the lattice constant is indicated by crosses.

In the design of the topological chain, the position of the photonic band gap and the interface mode are directly linked to design parameters such as the lattice period. This allows us to specifically target any wavelength within the gain spectrum of InGaAs purely by changing the design parameters. Tunability of the emission wavelength is demonstrated over the full telecom range, especially the technologically relevant O to C bands, in fig. 5 b. The demonstrated devices differ only in their lattice constant, the linear relation to the emission wavelength allows easily predicable tuning, as expected for a photonic crystal laser.

## DISCUSSION

Through the in-plane growth of InGaAs nanorods and their embedding in an SSH lattice, we realized the first co-planar hybrid active/passive photonic cavity and demonstrated single mode lasing at telecom wavelengths. The local co-placement of Si and III-Vs thereby allows us to selectively enhance the topological interface mode and achieve single mode lasing for any gain bandwidth. While this is already a significant result on its own, the underlying concept and fabrication platform is of high relevance for both the advancement of PICs and novel concepts in topological photonics. The combination of the active and passive material allows us to abstain from the introduction of loss into the photonic system, as previous efforts towards topological single mode lasers relied on[25,28,39]. Loss in these PT-symmetric hybrid gain/loss systems is typically introduced on top of existing gain material in the form of metals. Furthermore, this realization shows the potential of the hybrid III-V/Si concept as a versatile platform for realizing integrated topological photonic devices with enhanced functionalities. Compared to e.g. single mode microdisk lasers[25,37], the proposed one-dimensional array can be easily integrated into existing PICs through direct in-plane coupling to Si waveguides. The adjacent placement of the III-V material next to Si is crucial here, as it not only removes the necessity for large, complex coupling schemes[40–42] in between two vertically separated layers, but also opens new paradigms in designing photonic systems.

## METHODS

### Simulations

3D finite-difference time domain (FDTD) simulations of the ideal topological lattice were carried out in the commercially available software Ansys Lumerical FDTD. We assume the same refractive index of $n = 3.5$ for all nanorods, which holds true for Si and InGaAs in the wavelength range of interest. Note that material or shape differences will not significantly impact the results due to device symmetry. The optical modes supported by the device were excited with a short pulse of randomly placed dipole sources. Mode properties such as wavelength and quality ($Q$) factor were evaluated using an apodization window to filter out the initial excitation pulse and remove any simulation-depended artifacts. The quality factor of each mode is calculated from the decay of the field intensity over time.

### Fabrication:
### Template-assisted selective epitaxy (TASE)

An SOI wafer with a Si device layer of 220 nm thickness is patterned by standard HBr-based silicon dry etching in an inductively coupled plasma reactive ion etch (ICP-RIE). A layer of silicon dioxide ($SiO_2$) is deposited by atomic layer deposition (ALD), encapsulating the Si features. This oxide shell will serve as the template for transferring the Si shape onto the III-Vs later. To achieve this, an opening is etched into the $SiO_2$ layer on one side of the Si structure and the uncovered Si underneath is etched away selectively using tetramethylammonium hydroxide (TMAH). The now hollow oxide template remains, taking over the exact shape of the structure defined in Si before. Additionally, a small segment of Si is left standing at the end of the template to serve as nucleation site for the following MOCVD growth step, during which the III-V semiconductors will fill up the template. After the desired growth of the III-V segment the remaining Si seed can be removed selectively using $XeF_2$ gas. MOCVD growth of $In_xGa_{1-x}As$ used trimethylindium, trimethylgallium, and tertiarybutylarsine as precursors with a V/III ratio of 55 and In/(In+Ga) molar flow of 0.3 for a target composition of $x = 0.5$.

**Characterization:**
**Micro-photoluminescence measurements**

Fabricated devices are characterized by micro-photoluminescence (µPL) at 100 K. We use a pulsed supercontinuum pump laser, which delivers ps pulses at a repetition rate of 78 MHz and a wavelength of 1100 nm to ensure excitation above the bandgap of InGaAs. A 100x objective lens focuses the excitation beam to a spot size of about 1-2 µm and then also collects the emission from the sample. The PL spectrum is obtained using a grating spectrometer and a cooled InGaAs line array detector. For measuring the photon lifetimes, we correlate the counts on an InGaAs single photon detector with the excitation pulse by time-correlated single photon counting (TCSPC).

**Supplementary information and data availability**

Supplementary material is available below. All data supporting the findings of this study will be made available on Zenodo.

**ACKNOWLEDGEMENT**

This work was supported by the Swiss National Science Foundation (grant 188173) and the National Research Foundation of Korea (grants 2019K1A3A1A14064815, 2020R1I1A3071811, and 2022M3H3A1085772). The authors thank the Cleanroom Operations Team of the Binnig and Rohrer Nanotechnology Center (BRNC) for technical support. The authors declare no conflict of interest.

# Supplementary information

# Single mode laser in the telecom range by deterministic amplification of the topological interface mode

Markus Scherrer[1], Chang-Won Lee[2], Heinz Schmid[1], Kirsten E. Moselund[3,4,*]

[1] IBM Research Europe – Zurich, 8803 Rüschlikon, Switzerland

[2] Institute of Advanced Optics and Photonics, Hanbat National University, Daejeon, South Korea

[3] Laboratory of Nano and Quantum Technologies (LNQ), Paul Scherrer Institut (PSI), 5232 Villigen, Switzerland

[4] Integrated Nanoscale Photonics and Optoelectronics Laboratory (INPhO), EPFL, 1015 Lausanne, Switzerland

[*] Correspondence: kirsten.moselund@psi.ch, +41 56 310 34 15

# I. Comparison between topological interface and bulk modes

With fig. S1, we want to emphasize the difference in electromagnetic field intensity distribution between the topological interface (TI) mode and the 'trivial' photonic crystal modes. This is key to our design concept as it enables the deterministic amplification of the desired topological interface mode. Fig. S1 a gives a cross section of the local field intensity through the center of the device, where the confinement in the one dimensional array leads to the strongest value. The positions of the two unit cell sites *A* and *B* are highlighted in different color at the top and bottom, allowing to directly relate intensity and position.

For the TI mode, intensity maxima fall exactly on the positions of the *A* sublattice and the minima overlap with the *B* sublattice. For the two trivial modes, the intensity is more evenly distributed within both sublattices. Note that the mode with a higher frequency has its highest intensity in the space between two nanorods – it belongs to the so-called 'air band', whereas the lower frequency side of the photonic band gap makes up the 'material band'.

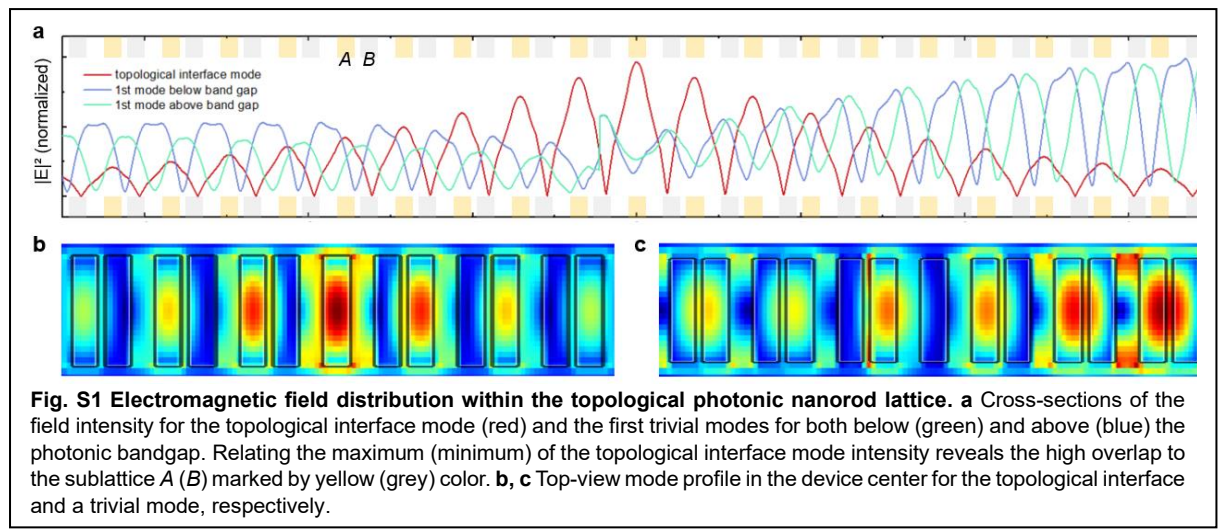

**Fig. S1 Electromagnetic field distribution within the topological photonic nanorod lattice. a** Cross-sections of the field intensity for the topological interface mode (red) and the first trivial modes for both below (green) and above (blue) the photonic bandgap. Relating the maximum (minimum) of the topological interface mode intensity reveals the high overlap to the sublattice *A* (*B*) marked by yellow (grey) color. **b, c** Top-view mode profile in the device center for the topological interface and a trivial mode, respectively.

The overlap fraction of each mode with the respective unit cell sites is given in table S1, again highlighting the difference of the TI mode for the two sublattice sites.

**Tab. S1 Calculated modal overlap the two sublattice sites.** The significant difference in distribution is obvious for the topological interface mode. Numbers in percentage, the remainder is distributed in the oxide cladding in between the nanorods.

|  | Sublattice *A* | Sublattice *B* |
|---|---|---|
| **Interface mode** | 47.7 | 12.5 |
| **'Air' mode** | 27.9 | 34.9 |
| **'Material' mode** | 31.8 | 37.2 |